\documentclass[9pt,conference]{IEEEtran}
\usepackage[preprint]{waspaa25}

\usepackage{booktabs}

\usepackage{bm} %
\usepackage[numbers,sort&compress]{natbib}
\setcitestyle{max=1}

\newcommand{\mixture}{y}
\newcommand{\source}{s}

\newcommand{\degradationfunc}{f}
\newcommand{\degradationset}{\mathcal{F}}
\newcommand{\prior}{p}

\title{Music Source Restoration}

\name{Yongyi Zang$^{1}$,
      Zheqi Dai$^{2, 3}$,
      Mark D. Plumbley$^{3}$,
      Qiuqiang Kong$^{2\dag}$\thanks{\dag Please address correspondance to Qiuqiang Kong (qqkong@ee.cuhk.edu.hk)}}
\address{$^{1}$Independent Researcher, Seattle, WA, USA \;
$^{2}$The Chinese University of Hong Kong, Hong Kong SAR, China \\
$^{3}$CVSSP, University of Surrey, Guildford, UK
}

\begin{document}

\maketitle

\begin{abstract}
We introduce Music Source Restoration (MSR), a novel task addressing the gap between idealized source separation and real-world music production. Current Music Source Separation (MSS) approaches assume mixtures are simple sums of sources, ignoring signal degradations employed during music production like equalization, compression, and reverb. MSR models mixtures as degraded sums of individually degraded sources, with the goal of recovering original, undegraded signals. Due to the lack of data for MSR, we present RawStems, a dataset annotation of 578 songs with unprocessed source signals organized into 8 primary and 17 secondary instrument groups, totaling 354.13 hours. To the best of our knowledge, RawStems is the first dataset that contains unprocessed music stems with hierarchical categories. We consider spectral filtering, dynamic range compression, harmonic distortion, reverb and lossy codec as possible degradations, and establish U-Former as a baseline method, demonstrating the feasibility of MSR on our dataset. We release the RawStems dataset annotations, degradation simulation pipeline, training code and pre-trained models to be publicly available. 
\end{abstract}

\section{Introduction}
\label{sec:intro}

\begin{quote}
\textit{``It has been said: The whole is more than the sum of its parts. It is more correct to say that the whole is something else than the sum of its parts.''} 

\hfill \textit{--- Kurt Koffka~\cite{koffka2013principles}}
\end{quote}

Music Source Separation (MSS)~\cite{stoter2019open, defossez2019music} aims to decompose musical mixture signals into their constituent instrument signals, assuming that mixtures are simple sums of these components. Recent deep learning approaches have achieved significant progress, with state-of-the-art methods reaching signal-to-distortion ratios (SDR) exceeding 10 dB~\cite{lu2024music, wang2023mel}. Current methods can be broadly categorized into time-domain approaches (such as WaveUNet \cite{stoller2018wave}, ConvTasNet \cite{luo2019conv}, and multiple resolution analysis \cite{nakamura2021time}) that directly process audio waveforms, and time-frequency domain approaches that first transform waveforms into spectrograms before applying various neural architectures including fully connected networks \cite{uhlich2017improving}, RNNs \cite{huang2015joint}, CNNs \cite{takahashi2018mmdenselstm}, U-Nets \cite{jansson2017singing, hennequin2020spleeter}, BSRNNs \cite{luo2022music}, and BS-Roformers \cite{lu2024music}. Hybrid approaches \cite{defossez2021hybrid, rouard2022hybrid} have also emerged, combining both domains for improved performance.

However, current MSS evaluation suffers from a fundamental limitation: it assumes an idealized scenario where musical sources are directly summed to form mixtures. This assumption fails to capture the complexity of real-world audio production. Real-world musical recordings undergo numerous transformations throughout the production chain~\cite{watkinson2013art, zolzer2022digital}, including: room acoustics that introduce reverberation and modal resonances~\cite{sabine1928acoustics, bottalico2020reproducibility}; recording equipment with inherent non-linearities and frequency response characteristics~\cite{corbett2020mic, fahed2022comparison}; mixing processes involving dynamic compression, equalization, and harmonic distortion~\cite{senior2018mixing}; mastering that further modifies frequency balance and dynamic range~\cite{shelvock2012audio}; and distribution platforms that apply lossy compression algorithms introducing additional frequency loss and artifacts~\cite{islam2022communication}. In music production, music is created by summing individually degraded instrument signals (See Sec.~\ref{sec:task-formulation} for details). 

To address the gap between idealized source separation and real-world music production, we introduce the \textbf{Music Source Restoration (MSR)} task. The MSR task aims to recover the original, undegraded source signals from the mixture. As the degradation processes result in information loss, MSR exhibits an inherent many-to-one mapping: the same degraded signal can be mapped to multiple original signals. 
For instance, a degraded signal obtained by applying a lowpass filter on the original signal with a cutoff frequency of 2 kHz can be mapped to several original signals with different timbres. 
To further constraint the solution space, we additionally restrict the recovered source signals to be close to the distribution of unprocessed instruments. We propose two metrics to evaluate for the MSR task: Structural Similarity Index Measure (SSIM)~\cite{wang2004image} on the mel-spectrogram as a perceptual similarity metric~\cite{namgyal2023you}, and Scale-Invariant Signal-to-Distortion Ratio (SI-SDR)~\cite{le2019sdr} as a waveform-level metric. 

To address the limitations of existing MSS datasets~\cite{rafii2019musdb18, pereira2023moisesdb} that only contain pre-processed signals, we present RawStems, a new dataset that provides access to the original, unprocessed source signals necessary for training and evaluating MSR systems, annotating 578 songs from the `Mixing Secrets' multitrack library~\cite{senior2018mixing}. With 8 first-level instrument groups, 17 second-level instrument groups, and a total runtime of 354.13 hours. To our knowledge, RawStems is the largest unprocessed hierarchical musical source dataset. We propose possible degradations into five categories: spectral filtering, dynamic range compression, harmonic distortion, reverb and lossy codec, and provide Python implementation to simulate them. By combining U-Net~\cite{stoller2018wave} and RoFormer~\cite{wang2023mel}, we propose U-Former as a baseline method for our task and dataset, and report its performance. We hope our work marks the first step towards source restoration models that can be more practically used in music production settings. We release RawStems at \url{https://huggingface.co/datasets/yongyizang/RawStems}, code at \url{https://github.com/yongyizang/music_source_restoration} and model checkpoints at \url{https://huggingface.co/yongyizang/MSR_UFormers}.

\section{The Music Source Restoration Task}
\label{sec:task-formulation}
\subsection{Task Definition}
\label{ssec:definition}
The MSR problem models the observed mixture $\mixture \in \mathbb{R}^{L}$ as a sum of degraded sources, where $ L$ is the number of samples of the signal. The goal of MSR is to recover the original, unprocessed forms $\source_1, \source_2, \dots, \source_n$ similar to direct recordings from a live performance in a professional recording studio from the mixture $ \mixture $. Formally, given a set of degradation functions $\degradationset$, the mixture is expressed as $\mixture = \sum_{i=1}^n \degradationfunc_i(\source_i)$, where each $\degradationfunc_i \in \degradationset$ is applied to source $\source_i$. As mentioned in Sec.~\ref{sec:intro}, due to the many-to-one mapping inherent in degradation processes, we introduce \emph{neutrality} as a constraint: we require the recovered sources to be consistent with priors $\prior_i$ representing distributions of original, unprocessed musical stems. MSR is thus framed as conditional generation: given the observed mixture $\mixture$, priors $\prior_i$, and possible degradation functions $\degradationset$, we aim to generate sources $\source_i$ such that:
\begin{itemize}
    \item the source is \emph{plausible}, indicating that there exist degradation functions $\{\degradationfunc_i\}_{i=1}^n$ from $\degradationset$ satisfying $\sum_{i=1}^n \degradationfunc_i(\source_i) = \mixture$
    \item the source is \emph{neutral}, indicating that each $\source_i$ belongs under its respective prior $\prior_i$ (i.e., $\source_i \sim \prior_i$).
    
\end{itemize}

\subsection{The Degradation Set $\degradationset$}
\label{ssec:degradation-set}
Referencing common audio processing techniques used in music production~\cite{senior2018mixing, izhaki2017mixing}, we group possible degradation functions in degradation set $\degradationset$ into five categories: spectral filtering, dynamic range compression, harmonic distortion, reverb and lossy codec. Additionally, we incorporate in $\degradationset$ gain functions $F(x) = ax, a \neq 0$ to account for differences in audio volume across processed samples.

\textbf{Spectral Filtering.} We consider filtering through two approaches: a 16-band graphic equalizer (EQ)~\cite{greiner1983design} utilizing fixed center frequencies ($25$, $40$, $63$, $100$, $160$, $250$, $400$, $630$, $1000$, $1600$, $2500$, $4000$, $6300$, $10000$, $16000$, $20000$ Hz) with random gains uniformly distributed between $-6$ and $+6$ dB, and a parametric EQ~\cite{massenburg1972parametric} employing $1$-$5$ randomly selected filter bands. Parametric filters include lowpass ($200$-$8000$ Hz), highpass ($20$-$2000$ Hz), bandpass/bandstop/peak ($100$-$8000$ Hz), lowshelf ($50$-$1000$ Hz), and highshelf ($1000$-$10000$ Hz), each with $Q$ factors ranging from $0.5$-$10$ and gains from $-6$ to $+6$ dB. Filter implementations include Butterworth (maximally flat), Chebyshev Type I ($0.1$-$3.0$ dB passband ripple), Chebyshev Type II ($20$-$60$ dB stopband attenuation), Elliptic ($0.1$-$3.0$ dB passband ripple, $20$-$60$ dB stopband attenuation), and Bessel (linear phase) approaches, with even-valued filter orders between $2$-$8$~\cite{smith2007introduction}. For graphic EQ, band-specific $Q$ values are calculated as $Q = f_{\text{center}}/(f_{\text{upper}} - f_{\text{lower}}) \times 2.5$ for peak filters, while parametric filter coefficients are computed directly from center frequency, $Q$, and gain parameters, with all filters implemented as second-order sections to maintain computational stability across the entire $20$ Hz - $20$ kHz spectrum. We implement all filters based on the signal processing library of SciPy~\cite{gommers2024scipy}.

\textbf{Dynamic Range Compression.} We consider two types of dynamic range compression algorithms: compressor~\cite{giannoulis2012digital} and limiter~\cite{mcnally1984dynamic}. The compressor processes audio signals with a configurable threshold $T_c \in [-20, 0]$ dB, ratio $R_c \in [1.5, 10.0]$, attack time $A_c \in [1, 10]$ ms, and release time $R_{c,\text{rel}} \in [50, 200]$ ms. When the input signal amplitude exceeds $T_c$, the compressor attenuates the signal according to ratio $R_c$, where output level change equals input level change divided by $R_c$. The attack parameter $A_c$ determines how quickly compression engages when threshold is crossed, while $R_{c,\text{rel}}$ controls how rapidly the compressor returns to unity gain when signal falls below threshold. The limiter functions as an extreme compressor with infinite ratio, configured with threshold ($T_l \in [-10, 0]$ dB) and release time ($R_{l,\text{rel}} \in [50, 200]$ ms). Any signal exceeding $T_l$ is immediately attenuated to maintain the threshold ceiling, with fixed instantaneous attack time and configurable $R_{l,\text{rel}}$ governing return-to-unity behavior. We use the implementation provided by Pedalboard~\cite{pedalboard}.

\textbf{Harmonic Distortion.} Diverse sources of harmonic distortion is added throughout the mixing process~\cite{izhaki2017mixing, senior2018mixing}. For simplicity, we consider only hyperbolic tangent distortion~\cite{schuck2016audio}, implemented with a configurable drive parameter $D \in [0, 5]$ dB, primarily introducing predominantly odd-order harmonics to the signal. Signal amplitude is first amplified by the drive amount, then passed through the \texttt{tanh} non-linearity, creating a compressed waveform with soft clipping behavior. This process simulates analog saturation found in hardware circuits~\cite{tarr2023development}. We use the implementation provided by Pedalboard~\cite{pedalboard}.

\textbf{Reverb.} While many reverb types exist in practice~\cite{mcguire2020mixing}, we use the FreeVerb~\cite{smith2010physical}, which uses four Schroeder allpasses in series and eight parallel Schroeder-Moorer filtered-feedback combfilters for each audio channel to simulate a reverb tail. The FreeVerb is parameterized by room size $R_s \in [0.1, 1.0]$ controlling decay time of the comb filters, damping $D \in [0.1, 1.0]$ determining high-frequency absorption in the feedback loops, wet level $W_l \in [0.1, 0.5]$ adjusting the ratio of processed to unprocessed signal, and stereo width $S_w \in [0.1, 1.0]$ controlling the phase differences between left and right channels.

\textbf{Codec.} Similar to~\cite{li2025apollo}, we consider only the MP3 codec, as it is one of the most popular lossy codec in practice~\cite{sripada2006mp3}. We use the varying bitrate encoding with a quality parameter $Q_{\text{vbr}} \in [1.0, 9.0]$, where lower values produce higher quality audio with increased bitrates and less perceptual artifacts. We use the implementation provided by Pedalboard~\cite{pedalboard}, which uses the LAME MP3 Encoder~\footnote{\url{https://lame.sourceforge.io/}}.

\begin{figure*}[h]
\begin{center}
\includegraphics[width=\textwidth]{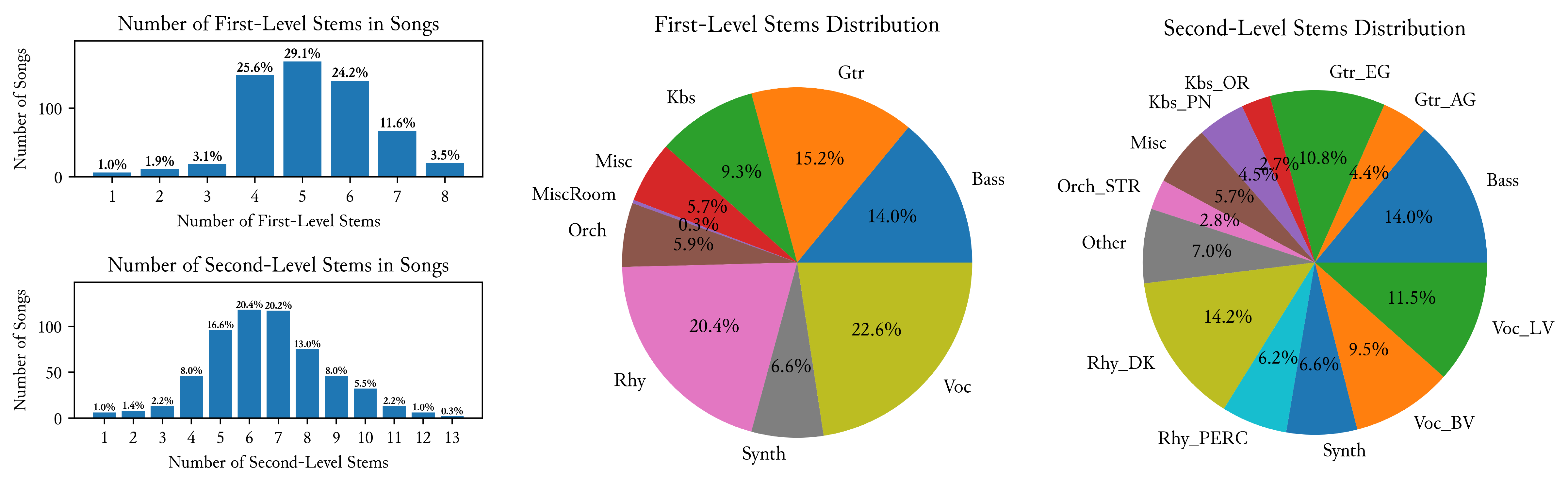}
\end{center}
\caption{Analysis on the RawStems dataset at first and second stem level. Stems distribution reported with active playing times.}
\label{fig:main-figure}
\end{figure*}

\subsection{Evaluation Metrics}
The MSR task requires evaluation of generated sources on both \emph{plausibility} and \emph{neutrality} (See Sec.~\ref{ssec:definition}). We employ SI-SDR and SSIM to assess these aspects respectively.

\textbf{SI-SDR.} We utilize Scale-Invariant Signal-to-Distortion Ratio (SI-SDR) to evaluate the \emph{plausibility} of generated sources. This metric is commonly used in Music Source Separation (MSS) tasks~\cite{luo2019conv} and is particularly appropriate in our context since gain functions are included in our degradation set $\degradationset$, making the scale-invariant property essential for fair evaluation.

\textbf{SSIM.} To assess \emph{neutrality}, we apply the Structural Similarity Index Measure (SSIM), which was originally developed to measure perceived similarity between original and degraded images~\cite{wang2004image} and has recently been adopted for audio perceptual similarity assessment~\cite{namgyal2023you}. We calculate power mel-spectrograms using 256 bins with a Hann window of 4096 samples and a hop size of 1024 samples to capture an appropriate level of detail.

Following standard MSS evaluation practices~\cite{stoter20182018}, we compute both SI-SDR and mel spectrogram SSIM on 10-second segments and report their means with 95\% confidence intervals.

\section{The RawStems Dataset}
The RawStems Dataset annotates 578 songs, annotated as 8 first-level instrument groups and 18 second-level instrument groups, with a total runtime of 354.13 hours. While the license of audio files belongs to the original creator, we provide the annotation in Apache 2.0 license.

\subsection{Instrument Taxonomy}
Instrument taxonomy, though well-established in research literature, presents significant challenges when designing a classification system that balances comprehensiveness with usability~\cite{pereira2023moisesdb}. Our taxonomy's primary level consists of eight major categories, informed by professional stem mastering practices, where eight stems represent the standard configuration for most summing machines (analog devices used for combining audio signals to achieve optimal analog-to-digital conversion). Each primary category is further divided into specific subcategories, each one with its unique code, yielding 17 subcategories:
\begin{itemize}
    \item Vocals (Voc): lead vocal (LV), backing vocal (BV), and group vocals (GV)
    \item Rhythm sections (Rhy): drum kits (DK) and percussions (PERC)
    \item Guitars (Gtr): acoustic (AG) and electric guitar (EG)
    \item Keyboards (Kbs): acoustic piano (PN), electric piano (EP), mellotrons (MTR), and organs (OR)
    \item Orchestra (Orch): strings (STR), brass (BR), and woodwinds (WW)
    \item Synthesizers (Synth): various electronic synthesizer sounds
    \item Bass: various bass sounds
    \item Misc: other signals (bells, effects, etc.)
    \item MiscRoom: Room microphones (capturing entire room ambience)
\end{itemize}

A significant challenge in instrument classification is addressing overlaps between categories. For instance, synthesizer strings could logically belong to either synthesizers or orchestral strings; similarly, synthesizer bass could be classified as either synthesizer or bass. To address these ambiguities, we established the following rules: all mellotron signals are categorized under keyboards/mellotrons regardless of the simulated instrument, and all synthesizer bass instruments are classified under bass rather than synthesizers. For edge cases not covered by these rules, classification decisions were made through best human subjective judgement, placing each sound in the most appropriate category based on timbral characteristics and musical function. For example, when a signal of a horn section contains both brass and woodwind instruments (such as trumpets and saxophones), we label it a brass in the orchestral class (Orch\_BR).

\subsection{Data Collection and Annotation}
Our annotation process followed three phases: (1) LLM-assisted auto-labeling using Claude 3.7 Sonnet to classify stems based on filenames into 18 subcategories or unknown (UNK) when uncertain; (2) manual verification of all songs containing ``UNK'' labels through human listening; and (3) manual verification of stems initially classified as ``Misc.'' to ensure proper categorization according to our taxonomy rules.

\subsection{Dataset Analysis}
\begin{figure}[t!]
\begin{center}
\includegraphics[width=0.9\linewidth]{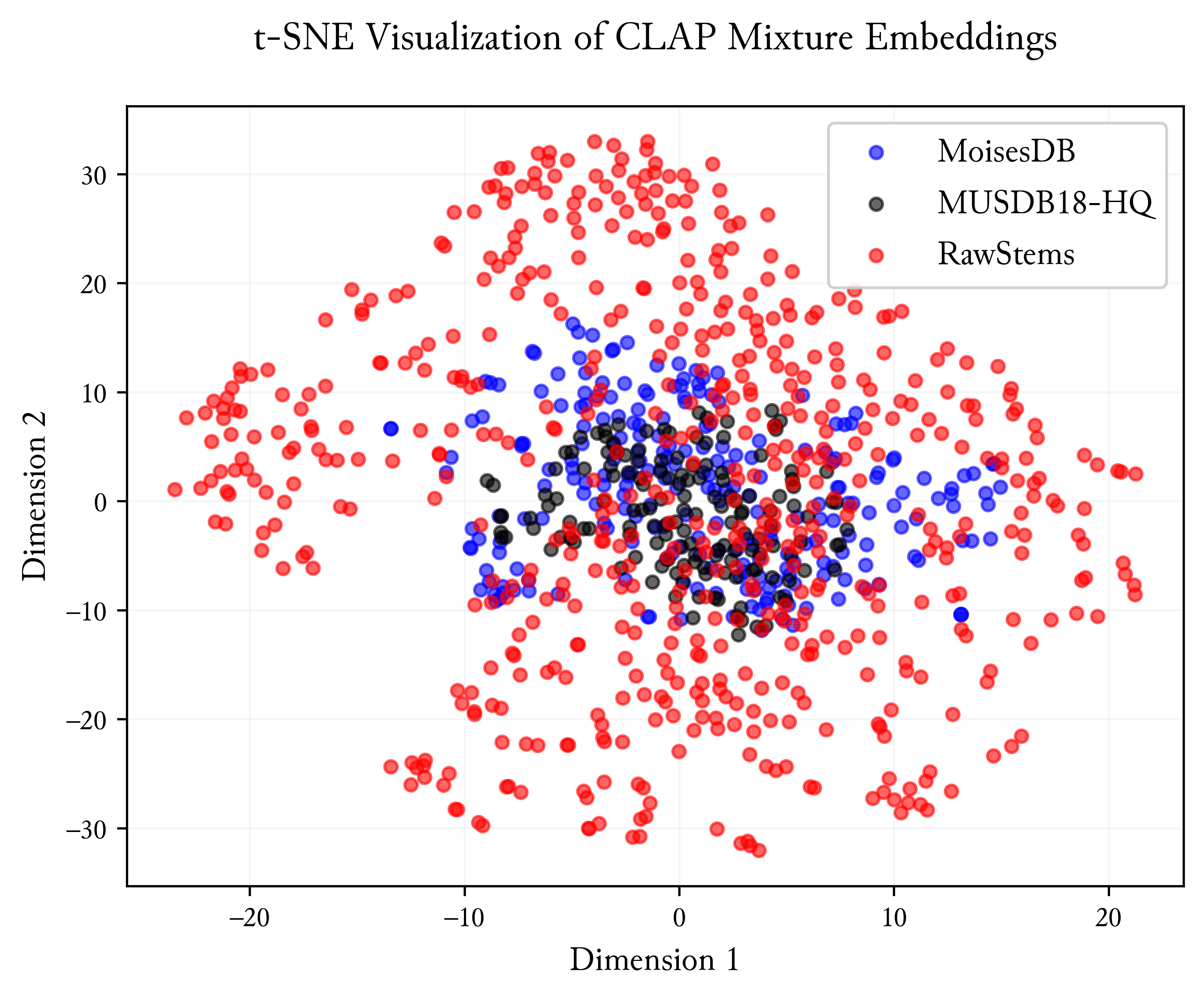}
\end{center}
\caption{t-SNE visualization for mixture-level CLAP embeddings of RawStems, MUSDB18-HQ and MoisesDB. Best viewed in color.}
\label{fig:mixture-level-clap}
\end{figure}

\begin{figure}[h]
\begin{center}
\includegraphics[width=\linewidth]{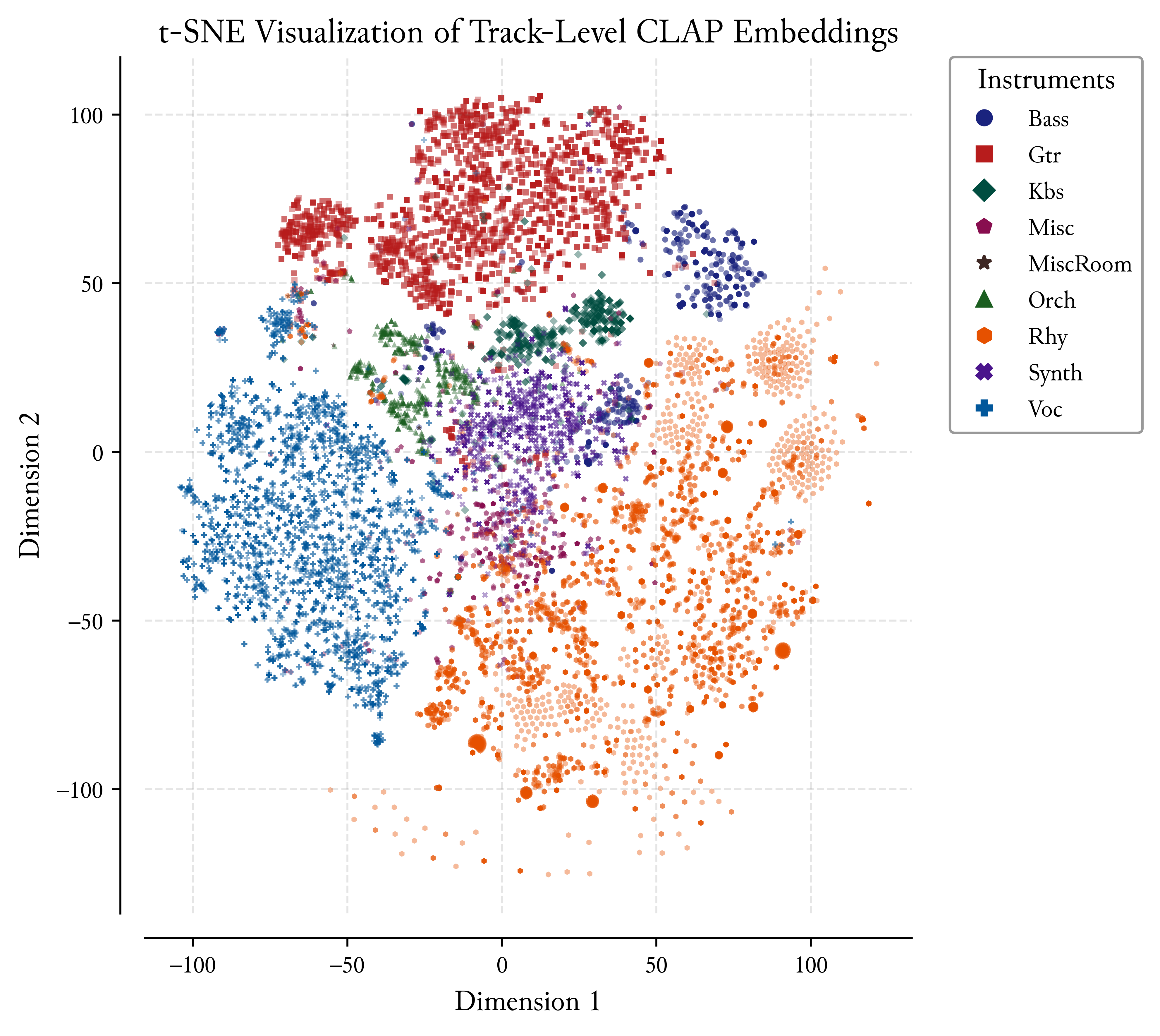}
\end{center}
\caption{t-SNE visualization for track-level CLAP embeddings of all individual audio files within RawStems. Best viewed in color.}
\label{fig:track-level-clap}
\end{figure}
An analysis of the stem-level statistics is presented in Figure~\ref{fig:main-figure}, revealing distinct patterns in instrumental composition across the dataset. The distribution shows a clear concentration, with 78.9\% of songs containing between 4-6 first-level stems, while 86.2\% feature 4-9 second-level stems. To count active playing time per instrument, we normalize each audio file, then sum the number of 1-second segments surpassing $-40$ dB RMS as the active playing time in seconds. our findings align with those reported by \cite{pereira2023moisesdb}, with vocals, guitars, bass, rhythm section, and keyboards collectively dominating the sonic landscape (81.5\% of total active playing time).

To gain a more comprehensive understanding of the hierarchical structure within RawStems, we computed CLAP~\cite{wu2023large} embeddings at both the individual track and mixture levels. Referencing~\cite{vosoughi2024learning}, we project the resulting embeddings into 2D space using t-SNE, as illustrated in Figure~\ref{fig:track-level-clap} for track-level analysis and Figure~\ref{fig:mixture-level-clap} for mixture-level analysis. At the track level, we observe clustering patterns that align with our expectations: tracks naturally segregate according to their first-level stem classifications, while forming subclusters that correspond to second-level stem categories or even more fine-grained instrument groupings. This hierarchical clustering validates the taxonomic organization of our dataset. When examining the mixture-level embeddings, RawStems demonstrates broader distribution in the embedding space compared to both MUSDB18-HQ and MoisesDB, indicating a more diverse range of musical compositions and timbral characteristics.

\begin{figure}[h]
\begin{center}
\includegraphics[width=\linewidth]{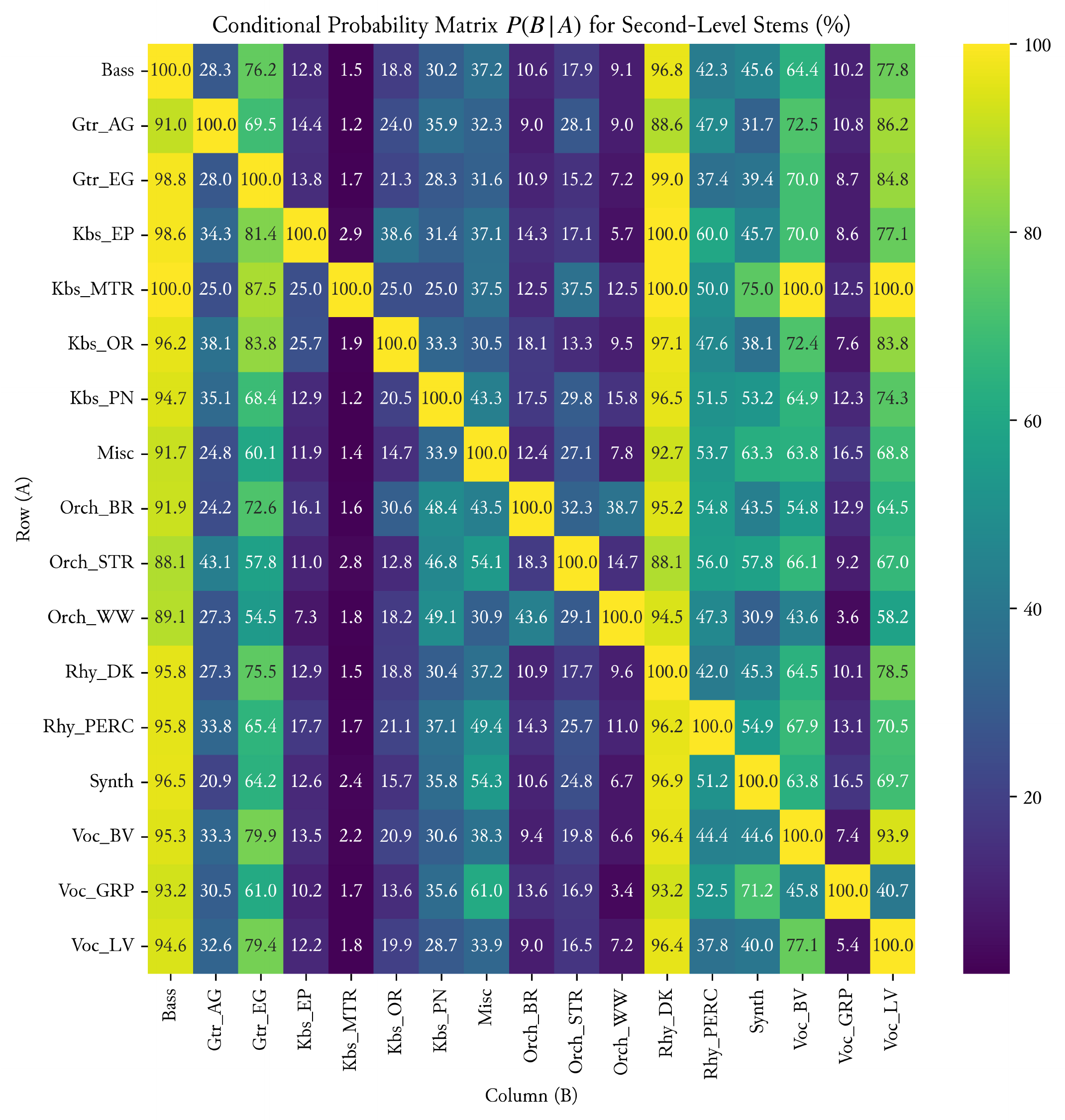}
\end{center}
\caption{Bayesian analysis for co-occurrence of instrument stems.}
\label{fig:bayesian}
\end{figure}

We also analyzed the co-occurrence patterns among different instrument stems by computing the conditional probability $P(B|A)$ (the probability of observing instrument stem $B$ given the presence of instrument stem $A$) focusing specifically on second-level instrument groupings. The results are visualized in Figure~\ref{fig:bayesian}. The analysis reveals that certain instruments, like bass, appear across nearly all songs. Additionally, we identified patterns consistent with real-world musical conventions, such as electric guitars frequently co-occur with both bass and acoustic guitars, while acoustic guitars rarely appear alongside the combination of bass and electric guitar.

\section{Baseline Experiments}
\begin{figure}[t!]
\begin{center}
\includegraphics[width=\linewidth]{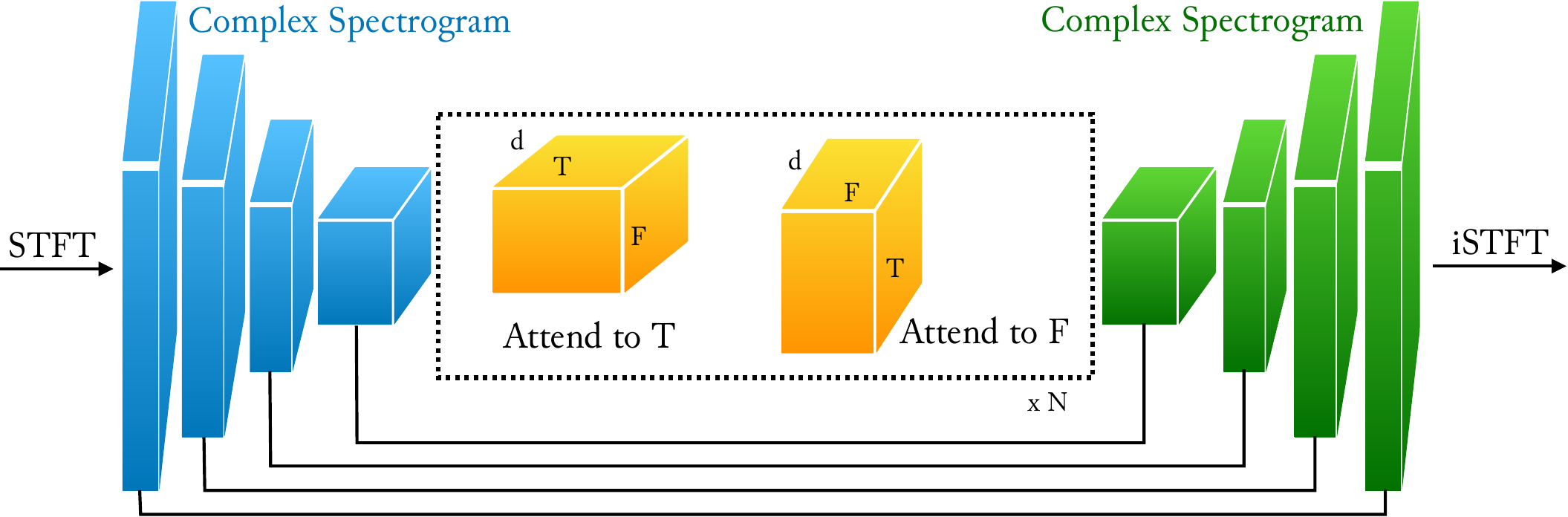}
\end{center}
\caption{Model architecture of U-Former.}
\label{fig:model-arch}
\end{figure}

We conduct a series of baseline experiments to demonstrate MSR feasibility and evaluate our proposed dataset. While ideal training would encompass all first- and second-level stems, this requires 25 distinct models—exceeding our compute budget constraints. We therefore train nine baseline models: seven representing first-level stem groups (excluding ``Misc.'') and two specifically for acoustic and electric guitar. Empirically, we observe that the U-Net architecture~\cite{hennequin2020spleeter, stoller2018wave} offers training efficiency but exhibits deficiencies in high-frequency detail reconstruction, whereas the BS-Roformer~\cite{lu2024music} delivers superior quality at significantly higher computational cost. To optimize this quality-efficiency trade-off, we propose U-Former (illustrated in Figure~\ref{fig:model-arch}), which incorporates RoFormer blocks within U-Net's bottleneck layer, following established hybrid approaches~\cite{braun2021towards, chen2022dtt, mezza2024benchmarking}. Our implementation utilizes four convolution blocks in both upsampling and downsampling paths with channel dimensions $[16, 64, 256, 256]$; the RoFormer component comprises 6 layers, each with 8 attention heads. Both input and output utilize complex STFT representations computed with 2048-sample FFT length and 441-sample hop size.

\subsection{Training Setup}
We train our model using a two-stage optimization strategy spanning $85000$ steps, which requires approximately $24$ hours on a single NVIDIA H20 GPU. The training process is carefully designed to balance reconstruction quality with perceptual fidelity through a progressive loss schedule. During training, we construct each batch by randomly sampling $10$-second audio segments from our dataset, a duration that provides sufficient temporal context while maintaining computational efficiency. We employ the Adam optimizer with a linear learning rate warmup over the first $1000$ steps to ensure stable initial convergence, and apply gradient clipping with a threshold of $2.0$ to prevent training instabilities.

The initial $50000$ optimization steps focus exclusively on accurate waveform reconstruction. During this stage, we use a learning rate of $5 \times 10^{-4}$ with a batch size of $16$ samples, optimizing only the waveform-level $L_1$ reconstruction loss. This reconstruction-only phase allows the model to learn fundamental audio representations without the complexity of adversarial training, establishing a strong foundation for subsequent perceptual refinement. For the final $35000$ steps, we introduce adversarial training following the HiFi-GAN framework~\cite{kumar2023high} to enhance perceptual quality. Due to the increased memory requirements from incorporating the discriminator, we reduce the batch size to $4$ samples while maintaining the generator learning rate at $5 \times 10^{-4}$ and setting the discriminator learning rate to $5 \times 10^{-5}$, one order of magnitude lower for training stability.

The combined loss function during the adversarial training stage is formulated as 
\begin{equation}
    \mathcal{L}_{\text{total}} = \mathcal{L}_{\text{recon}} + \lambda_g \mathcal{L}_{\text{gen}} + \lambda_f \mathcal{L}_{\text{feat}}
\end{equation}
where $\mathcal{L}_{\text{recon}}$ is the $L_1$ reconstruction loss, $\mathcal{L}_{\text{gen}}$ is the generator adversarial loss, and $\mathcal{L}_{\text{feat}}$ is the feature matching loss. We set the loss weights to $\lambda_g = 0.001$ and $\lambda_f = 0.01$, empirically chosen to balance reconstruction accuracy with perceptual quality. The discriminator architecture follows the multi-scale discriminator design, which evaluates audio at different temporal resolutions. The generator, feature-matching and discriminator losses are computed as 
\begin{equation}
    \mathcal{L}_{\text{gen}} = \mathbb{E}_{x \sim p_{\text{data}}} \left[ (1 - D(G(x)))^2 \right]
\end{equation}
\begin{equation}
    \mathcal{L}_{\text{feat}} = \mathbb{E}_{x \sim p_{\text{data}}} \left[ \sum_{l=1}^{L} \frac{1}{N_l} \| D^{(l)}(x) - D^{(l)}(G(x)) \|_1 \right]
\end{equation}
\begin{equation}
\mathcal{L}_D = \mathbb{E}_{x \sim p_{\text{data}}} \left[ \sum_{l=1}^{L} \left[ D^{(l)}(G(x))^2 + (1 - D^{(l)}(x))^2 \right] \right]
\end{equation}
where $G(\cdot)$ is the generator (the restoration network in our case), $D^{(l)}$ represents the $l$-th layer features of the discriminator, and $N_l$ is the number of elements in the $l$-th layer. This two-stage training strategy effectively balances computational efficiency with model performance, allowing the network to first master accurate reconstruction before learning to generate perceptually convincing audio through adversarial training.

\subsection{Degradation Setting}
To create robust training examples, we independently aggregate randomly sampled audio signals from within the target stem group and from all sources outside the target group. For target stems, we apply a suite of degradation effects (equalizer, downsampling, compression, distortion, and reverb) each with 50\% probability of application. Similarly, for the mixture signals, we employ limiter, resampling, and codec artifacts, also each with 50\% probability. When any degradation type is selected, we uniformly sample across its parameter space to ensure diverse transformations. To further enhance robustness against noise, we augment 50\% of mixtures with Gaussian noise featuring means uniformly distributed between 0.001 and 0.005. The final mixing process combines target stems with background signals at signal-to-noise ratios ranging from -5 to 20 dB.

\subsection{Datasets}
For each stem group in RawStems, we randomly allocate songs containing the target stem to the testing set by selecting either 10 songs or 10\% of the total songs for that stem (whichever is larger). All remaining samples are assigned to the training set. To generate a comprehensive evaluation benchmark, we process the testing set through our degradation pipeline $1000$ times, creating a diverse array of degradation scenarios. To facilitate reproducibility, we make this offline-generated testing set publicly available alongside our implementation.

\begin{table}[t!]
\centering
\caption{Evaluation results for baseline experiments.}
\label{tab:baseline-results}
\begin{tabular}{lcc}
\toprule
\textbf{Instrument/Group} & \textbf{SSIM$_\text{Mel}$ $\uparrow$} & \textbf{SI-SDR (dB) $\uparrow$} \\
\midrule
\textbf{Vocals} & 0.5058 ± 0.0089 & 1.86 ± 0.77 \\
\textbf{Keyboards} & 0.4941 ± 0.0094 & -2.75 ± 0.92 \\
\textbf{Synth} & 0.4004 ± 0.0102 & -5.65 ± 0.95 \\
\textbf{Guitars} & 0.5564 ± 0.0098 & -2.40 ± 0.83 \\
\textbf{Rhythm} & 0.5927 ± 0.0087 & -0.11 ± 0.88 \\
\textbf{Bass} & 0.4437 ± 0.0086 & -7.90 ± 1.29 \\
\textbf{Orchestra} & 0.4722 ± 0.0103 & -2.37 ± 0.92 \\
\textbf{Guitars - Acoustic} & 0.5543 ± 0.0099& -1.14 ± 0.87\\
\textbf{Guitars - Electric} & 0.5875 ± 0.0097& -1.68 ± 0.81\\
\midrule
\textbf{VoiceFixer~\cite{liu2021voicefixer}} & 0.2543 ± 0.0052 & -40.72 ± 0.87\\
\textbf{Guitars - Acoustic (+30k steps)} & 0.5619 ± 0.0096 & -1.31 ± 0.88\\
\textbf{Guitars - Electric (+30k steps)} & 0.6019 ± 0.0093 & -1.25 ± 0.76\\
\bottomrule
 & &\\
\end{tabular}
\end{table}

\subsection{Evaluation}
We show evaluation results in Table~\ref{tab:baseline-results}. As expected, we observe that the SSIM metric and SI-SDR metric only loosely correlate with each other, suggesting they evaluate different aspects of the restored audio samples. 
Amongst which, we notice that vocal performance is best while bass performance is worst, suggesting for different instrument class the difficulty of the task may vary, as similarly observed for the MSS task~\cite{stoter2019open}. Interestingly, the acoustic and electric guitar model both behaves better than the general guitar model, suggesting benefits for training fine-grained models.

Notably, SI-SDR values are substantially lower than those typically observed in both MSS and audio restoration tasks~\cite{godsill2002digital}. We investigated three potential explanations: (1) flaws in our model architecture, training pipeline, or dataset; (2) insufficient training convergence; or (3) inherent task complexity. To test hypothesis (1), we evaluated a pre-trained VoiceFixer~\cite{liu2021voicefixer} model on our vocal test set, as it is trained to recover from both noise, reverb, low frequency rate and clipping, and thus is closest to the MSR task definition. We found it performed substantially worse than our model (SI-SDR near -40 dB). Manual inspection revealed its inability to properly separate vocal content from background instruments and reproduce high-frequency vocal details. For hypothesis (2), we extended training for acoustic and electric guitar models by 30,000 additional steps, yielding minimal improvement, indicating convergence had already occurred. These findings support hypothesis (3)—that MSR is inherently more challenging than related audio tasks—underscoring the need for dedicated research to develop more effective MSR models.

\section{Discussion}
\label{sec:discussion}

Our baseline results reveal an unusual gap between waveform‐level and perceptual metrics: SI-SDR values remain modest (often negative) despite \textit{mel}-SSIM scores that are on par with—or better than—typical music-source–separation baselines.  This confirms that MSR is intrinsically harder than MSS: each degraded mixture stems from an \emph{unknown} sequence of nonlinear, many-to-one transformations, so perfectly reconstructing phase and amplitude is neither feasible nor strictly necessary for musical usefulness.  Future work should therefore explore evaluation schemes that combine perceptual similarity with task-oriented objectives (e.g.\ remixing or effect-re-application), as well as crowd-sourced mean opinion score (MOS) studies, to complement SI-SDR.

A second limitation lies in the \emph{synthetic} nature of our degradation pipeline.  Although the chosen effects (EQ, compression, saturation, reverberation, lossy codecs) cover the bulk of practical production steps, their parameter distributions were selected heuristically.  Real-world sessions exhibit correlations across effects (e.g.\ compression followed by EQ boosts) and producer-specific preferences that our independent sampling ignores.  Extending RawStems with \emph{profiling} of commercial stems, or weakly supervised learning on user-generated stems scraped from multitrack communities would yield more realistic mixtures and stronger generalization tests.

Finally, our single-stem training regime highlights scalability challenges: computing budgets forced us to train nine separate U-Former models, yet preliminary experiments on narrower sub-classes (acoustic vs.\ electric guitar) improved both metrics.  These findings suggest that (i) \emph{multi-task} or \emph{Mixture-of-Experts} architectures could share low-level representations while retaining class-specific experts, and (ii) hierarchical prompts (e.g.\ “guitar:acoustic”)~\cite{liu2022separate} may let a single conditional model handle the full stem taxonomy.  We release our code and data to encourage exploration along these directions.

In all, we call for attention from the research community to tackle this challenge together towards building useful creative tools.

\section{Conclusions}
We introduced Music Source Restoration (MSR), a novel task addressing the gap between idealized source separation and real-world audio processing needs by modeling complex degradations in music production. We propose RawStems dataset, the largest-scale dataset for musical sources, annotating 578 songs with hierarchical instrument annotations across 8 primary and 17 secondary categories. We validate the feasibility of task and RawStems through baseline models. To our knowledge, the RawStems dataset is the first publicly available dataset that can be applied to address the MSR problem. In future, we plan to restore large-scale audio recordings from the internet to construct music stems dataset.

\clearpage
\bibliographystyle{IEEEtran}
\bibliography{refs25}
\end{document}